%% file: autoScale_guide.tex
\documentclass[12pt]{article}

\usepackage[a4paper]{geometry}
\usepackage{graphicx}
\usepackage{amsmath}
\usepackage{amssymb}
\usepackage{times}
\usepackage[pdftex]{color}
\usepackage{listings}

\definecolor{light}{gray}{.85} 
\usepackage[pdftex, draft=false, colorlinks=false, pdfborder={0 0 0}]{hyperref}

\newcommand{\myProg}{{\tt autoScale.py}}
\newcommand{\myVec}[1]{{\underline #1 }}
\long\def\symbolfootnote[#1]#2{\begingroup%
\def\thefootnote{\fnsymbol{footnote}}\footnote[#1]{#2}\endgroup}

\newcommand{\bibname}{Bib}

\begin{document}
\begin{titlepage}
\begin{center}
{\LARGE \myProg{} -- A program for automatic finite-size scaling analyses: A user's guide}
\vskip 0.5cm
O.~Melchert\symbolfootnote[1]{
E--mail:
\mbox{{\tt melchert@theorie.physik.uni-oldenburg.de}}}\\
\begin{small}
\em{Institut f\"ur Physik, Carl-von-Ossietzky Universit\"at Oldenburg,\\ 
26111 Oldenburg, Germany}\\
\end{small}
\end{center}
\vskip 1cm
\begin{abstract}
\myProg{} is a program that performs an automatic finite-size scaling analysis for
given sets of simulated data. It implements a quite general scaling assumption
and optimizes an initial set of scaling parameters that enforce a data collapse 
of the different data sets. The presented guide describes how the program works,
it presents a detailed example and finally gives some hints on how to
improve the results of a scaling analysis.
\end{abstract}
\vskip 1cm
\tableofcontents
\end{titlepage}
\include{autoScale_main}

\end{document}

%% file: autoScale_main.tex
%
%
%
\section{What is \myProg{}?}
\label{ASAsect01}
\myProg{} \cite{ASAmyProg} is a python \cite{ASApyRef} implementation of a program that performs an automatic 
finite-size scaling (FSS) analysis. 
More precise, \myProg{} uses \emph{data collapse techniques} so as to quantify 
the property of scaling.

Upon introduction of properly scaled variables, \myProg{} extracts numerical 
values for critical points and critical exponents that describe a given 
critical phenomenon. 
For the purpose of illustration consider
a quantity $y(x,L)$ that depends on two variables, $x$ and $L$. 
The precise meaning of the quantity $y$ depends on the actual problem.
E.g., in connection to percolation it may describe the relative size of the largest 
cluster of occupied sites. Practically it can be any observable for which one aims 
to extract the critical behavior.
The variables $x$ and $L$ are also problem dependent and represent adjustable 
parameters, e.g., in the problem of random percolation, these are the 
occupation probability and the system size.
Now, the scaling assumption states that if the above quantity is considered to obey 
scaling, it can be expressed as
\begin{eqnarray}
y(x,L)= L^{-b}~f[(x-x_c) L^a], \label{eq:scalingAssumption}
\end{eqnarray}
wherein $a$ and $b$ are dimensionless critical exponents, $x_c$ is called critical point 
and $f[ \cdot ]$ is a scaling function.   
To support intuition: if $L$ plays the role of a length scale, then the exponent $a$ 
measures the ``scaling dimension'' of the observable $x$. Likewise, the exponent $b$ 
measures the scaling dimension of the quantity $y$.

According to Eq.\ \ref{eq:scalingAssumption}, data curves of $y(x,L)$ for 
different values of $x$ and $L$ fall on top of each other if $y(x,L) L^{b}$ is plotted
against the combined quantity $u\!\equiv\!(x\!-\!x_c) L^a$ and if the free parameters $a$ and $b$ of 
the scaling assumption are chosen properly.    
Therein, the latter quantity $u$ corresponds to a polynomial 
approximation of the scaling function $f[u]$ up to lowest order in $u$. This
is certainly valid close to the critical point $x_c$, where $u\!\ll\!1$, and it usually
suffices to get reliable estimates for critical points and exponents.
Hence, to put it more procedural, one can estimate critical exponents quite precise if 
one employs the scaling assumption above, starts with a rough guess for the numerical 
values of the scaling parameters therein
and adjusts them so as to yield a satisfactory data collapse.

If one tries to characterize a critical phenomenon by means of computer simulations, such 
a scaling analysis is indispensable in order to infer critical points/exponents
from the simulation data.
Since the reliability of the critical properties thus obtained depends 
on the quality of the corresponding data collapse, it is desirable to 
perform the scaling analysis with the assistance of a computer. One
should not leave this crucial task to the competence of the bare eye.
Moreover, an automatic scaling analysis performed by algorithmic means,
ensures that the data collapse is attained in a systematic and
reproducible manner.
Up to now there is no ``ultimate'' method that yields an optimal data 
collapse, and within the community of computational experimentalists
there is --up to now-- no consensus of how to perform a proper finite-size 
scaling analysis. Instead, there are several methods 
\cite{ASAhoudayer2004,ASAbhattacharjee2001,ASAkawashima1993,ASAwenzel2008,ASAwinter2008} that find 
application once in a while. 
Here, \myProg{} implements the method described in Ref.\ \cite{ASAhoudayer2004}.
Until now, \myProg{} has successfully been used to perform the scaling analyses in 
Refs.\ \cite{ASAmelchert2008,ASAapolo2009,ASAmelchert2009,ASAwolfsheimer2009}.

\section{Program description}
\label{ASAsect02}

As pointed out above, \myProg{} can be used as a black box for performing automatic FSS analyses. 
In order to benefit from \myProg{} you will first need some unscaled data (see subsection \ref{ASAsect02subsect01}), 
that you wish to postprocess via a FSS analysis. 
Just supply the raw data and an ``educated guess'' for the appropriate scaling parameters
that yield a data collapse, and \myProg{} will take care of data scaling and scaling--parameter optimization. 

The quality of the data collapse, i.e.\ a measure of how well the scaling assumption fits the given data, 
is measured by the mean square distance of the data sets to the unknown master curve in units of the 
standard error \cite{ASAhoudayer2004}.
Upon execution of \myProg{}, the quality (or more precise: the scaling parameters that determine the quality)
is optimized via the downhill simplex algorithm of Nelder and Mead (see \cite{ASAnum_rec}, Chapter 10.4). 

The algorithmic procedure used to obtain the data collapse employs repeated application of linear 
least-squares methods \cite{ASAnum_rec} and goodness-of-fit tests \cite{ASAnum_rec,ASApractical_guide}.
Finally, \myProg{} provides numerical values for the scaling parameters, error estimates for the 
scaling parameters and a measure for the quality of the respective data collapse.

In the following, subsection \ref{ASAsect02subsect01} will specify the format of the input data that is 
passed to \myProg{} and subsection \ref{ASAsect02subsect02} it will explain how \myProg{} works.   
Finally, subsection \ref{ASAsect02subsect03} briefly outlines the general idea of the downhill simplex method.

\subsection{Required format of the input data} 
\label{ASAsect02subsect01}
As a precondition for \myProg{} to work, you will need some input data in a specific format.
If you are familiar with the python programming language, you can easily extend \myProg{}
by a function similar to the class method {\tt myRawData.fetchData()}, that reads data in 
your favored data format.
The standard format accepted by \myProg{} is described in the following. 

A data file is expected to contain three columns of numbers, there may be more but only the first three 
are considered. These columns are expected to list the $x$-values, $y$-values and the corresponding 
standard errors $dy$ in the form:

\begin{lstlisting}
# data file name: orderParam_L512.dat 
# <x-Values>  <y-Values>   <dy-Values>
0.565000000   0.019801567  0.000058048 
0.566750000   0.022313848  0.000068007 
0.568500000   0.025335819  0.000079926 
0.570250000   0.029096301  0.000095645 
0.572000000   0.033696019  0.000114802 
...           ...          ...
\end{lstlisting}
Note that \myProg{} treats lines starting with the symbol \verb'#' as a comment and it disregards empty lines.
Suppose you have produced data in the above format for several system sizes, 
say $L = 32$, $64$, $128$, $256$ and $512$, for which you aim to perform a FSS analysis. 

Create a configuration file, e.g.\ {\tt inputFiles.dat}, that lists the paths 
to the data files to be scaled along with the corresponding system sizes. 
For instance:

\begin{lstlisting}
# configuration file name: inputFiles.dat
# 	<path to file>               <system size L>
./ORDER_PARAMETER/orderParam_L32.dat     32
./ORDER_PARAMETER/orderParam_L64.dat     64
./ORDER_PARAMETER/orderParam_L128.dat    128
./ORDER_PARAMETER/orderParam_L256.dat    256
./ORDER_PARAMETER/orderParam_L512.dat    512
\end{lstlisting}
and you're almost done. \myProg{} will read this configuration file and accumulate 
the raw data from the specified data files to perform a scaling analysis.

\subsection{Invoking \myProg{}} 
\label{ASAsect02subsect02}
In advance, you should have an idea what scaling laws will fit your needs, 
and at least a rough guess regarding the values of the scaling parameters therein. 
Currently, \myProg{} implements the following scaling transformation:  
\begin{eqnarray*}
x  &&\mapsto \quad (x-x_c) \cdot L^a \\
y  &&\mapsto \quad \phantom{d}y\cdot L^b \\
dy &&\mapsto \quad dy\cdot L^b. 
\end{eqnarray*}
If there are different scaling laws needed to fit your observables, 
and if you are familiar with the python programming language, you can easily 
extend \myProg{} by implementing a scaling assumption similar to the class 
\verb'myScaleAssumption' that properly fits your requirements.

Invoking \myProg{} with the option \verb'-help' as \verb|python autoScale.py -help| yields information on how to use the program:

\begin{lstlisting}
melchert@comphy03:~/PYTHON/FSS_EXAMPLE> python autoScale.py -help

NAME
        autoScale.py -- a program for automated finite size scaling analyses

SYNTAX
        python autoScale.py -f inFile [-o outFile] [-xc val, -a val, -b val] 
                                [-xr val val] [-showS] [-getError]

        python autoScale.py [-help] [-version]

OPTIONS
        -help                    -- write usage and exit program
        -version                 -- write version number and exit
        -f   <inFile>            -- configuration file containing list of 
                                    paths to data files
        -o   <outFile>           -- path to output file (default: stdout)
        -xc  <float>             -- estimate for critical point (default: 0.0)
                                    if called as '-xc!', <float> is fixed during 
                                    parameter optimization 
        -a   <float>             -- estimate of exponent a (default: 0.0)
                                    if called as '-a!', <float> is fixed during 
                                    parameter optimization 
        -b   <float>             -- estimate of exponent b (default: 0.0)
                                    if called as '-b!', <float> is fixed during 
                                    parameter optimization 
        -xr <float> <float>      -- lower/upper boundary of interval on 
                                    rescaled x-axis for which scaling 
                                    analysis should be performed
        -showS                   -- report quality 'S' during minimization 
                                    procedure
        -getError                -- compute errors for scaling parameters 
                                    using S+1 analysis

EXAMPLE
        python autoScale.py -f dataFiles.dat -xc 0.592541 -a 0.754524 -b 0.107421
                      -xr -1. 1. -getError -o test.out -showS

\end{lstlisting}

\subsection{Minimization of the objective function}
\label{ASAsect02subsect03}
For the minimization of the objective function $S$, i.e.\ the optimization of the 
scaling parameters $\myVec{x}=(x_c,a,b)$, \myProg{} uses the downhill simplex algorithm of
Nelder and Mead \cite{ASAnelder1965}.
The simplex algorithm has proved to be a reliable tool, if one deals with unconstrained 
multidimensional optimization. 

In general, the task of the simplex algorithm (SA) is to minimize the value of a scalar 
function $S(\myVec{x})$ with $\myVec{x}\in \mathbb{R}^{n}$ that depends on $n$ variables. 
To do so, the SA requires only function evaluations and no function derivatives, hence it 
belongs to the so called direct search methods. 
The SA algorithm maintains a simplex that consists of $n+1$ vertices $\myVec{x}^{(1)},\ldots,\myVec{x}^{(n+1)}$, 
enclosing a finite $n$--dimensional volume, along with the corresponding function values 
$S^{(i)}\equiv S(\myVec{x}^{(i)})$. 

As a prerequisite, one has to provide an initial simplex with $n+1$ trial points $\myVec{x}_{{\rm ini}}^{(i)}$. 
The SA will iteratively improve the objective function values $S^{(i)}$ towards a local minimum value,
until it reaches a predefined minimal extension. 
For the improvement of the function values, the algorithm uses four different elementary 
manipulations of the simplex.
These are reflection, expansion, contraction and shrinkage. 
These manipulations govern the performance of the SA. 

A typical iteration step starts by ordering the $n+1$ vertices according to their function values $S^{(i)}$ 
in non-decreasing order. 
The vertex with the lowest function value ($S^{(1)}$), i.e.\ $\myVec{x}^{(1)}$, is referred to as \emph{best} point. 
The vertex with the largest function value ($S^{(n+1)}$), i.e.\ $\myVec{x}^{(n+1)}$, marks the \emph{worst} point. 
During one iteration, the algorithm tries to get rid of the worst function value using a sequence 
of geometrical transformations using the elementary manipulations mentioned above. 

\section{Example of a scaling analysis using \myProg{}}
\label{ASAsect03}
Within this section, the use of \myProg{} is demonstrated using data simulated for the model 
of $2d$ random site percolation \cite{ASAstauffer1979}.
The raw data and scripts used below are contained in the supplementary
archive \verb'SCALING_ANALYSIS.tar.gz'.
As observable, the order parameter of the percolation model, i.e.\ the relative size of
the largest cluster of occupied nearest neighbor sites
$P_\infty$ 
is considered.
Related to this observable, the scaling assumption takes the form 
\begin{eqnarray}
P_\infty(p,L)=L^{-\beta/\nu} f[(p\!-\!p_c)L^{1/\nu}], \label{eq:orderParam}
\end{eqnarray} 
where the parameter $L$ signifies the lateral extension of the square lattice, i.e.\ the system size, 
and the parameter $p$ indicates the occupation probability for a site on the lattice. 
Note that Eq.\ \ref{eq:orderParam} uses the standard notation of percolation theory.
Therein, the usual critical exponents are $\nu$ and $\beta$, where 
$\nu$ describes the divergence of a typical length scale as one approaches the 
critical point $p_c$ and $\beta$ is called the order parameter exponent.
In terms of the scaling form presented in Eq.\ \ref{eq:scalingAssumption}, one then has
$x=p$, $y(x,L)=P_\infty(p,L)$, $a=1/\nu$ and $b=\beta/\nu$.

For the scaling analysis, square systems of side length $L=128$, $256$, and $512$ are considered.
First, the optimal scaling parameters are determined in subsection \ref{ASAsect03subsect01}, 
then the errors of the scaling parameters are estimated in subsection \ref{ASAsect03subsect02}, and
the scaling plot that illustrates the data collapse is presented in subsection \ref{ASAsect03subsect03}. 

\subsection{Determine the optimal scaling parameters}
\label{ASAsect03subsect01}
If one settles for an approximation of the scaling function up to lowest order in 
the respective argument, as mentioned above in section \ref{ASAsect01}, one can expect that the scaling assumption 
holds best close to the critical point and that there are corrections to the scaling 
behavior outside the critical scaling window. In this regard, it might be useful to 
restrict the scaling analysis to a finite interval on the rescaled abscissa.
By trend, if one performs the scaling analysis for an interval 
$\Delta x$ of decreasing width that contains the critical point, 
then the quality of the data collapse improves. However, one should always find
a fair compromise between a small value of $S$ and a reasonably large interval 
$\Delta x$ on the rescaled abscissa.

To get a grip on how sensitive the scaling parameters are with respect to the interval
$\Delta x$, one can perform a repeated scaling analysis by using, e.g., a shell script
similar to the following

\begin{lstlisting}
# FILE:  scalingScript.sh
# TASK:  invoke scaling analysis program for different intervals xr
# USAGE: bash scalingScript.sh
for MAX in  1.5 1.25 1.0 0.75 ; 
do
 for MIN in -2.25 -2.0 -1.75 -1.5 -1.25 -1. -0.75 ;
 do
  python autoScale.py -f inputFiles.dat -o scaled_L512_256_128.out \
  	        -xc 0.5927 -a 0.75 -b 0.104 -showS -xr ${MIN} ${MAX}
 done
done
\end{lstlisting}
The output file \verb'scaled_L512_256_128.out' then contains the results of the
scaling analyses in the form:

\begin{lstlisting}
dx = [-2.250000:1.500000]  xc = 0.592571  a = 0.753171  b = 0.106964  S = 1.993244
dx = [-2.250000:1.250000]  xc = 0.592619  a = 0.752468  b = 0.106216  S = 1.505830
dx = [-1.750000:1.000000]  xc = 0.592687  a = 0.749537  b = 0.105055  S = 1.153121
dx = [-1.500000:1.000000]  xc = 0.592708  a = 0.747566  b = 0.104459  S = 1.076691
dx = [-1.000000:1.000000]  xc = 0.592686  a = 0.748386  b = 0.104841  S = 0.735622
dx = [-1.000000:0.750000]  xc = 0.592689  a = 0.749151  b = 0.104670  S = 0.767656
dx = [-0.750000:0.750000]  xc = 0.592709  a = 0.746708  b = 0.104098  S = 0.769959
\end{lstlisting}
Note that only a few lines of the output file are shown.

According to the scaling analysis, the best data collapse ($S\approx0.74$) was obtained for the
interval $\Delta x \in [-1.0:1.0]$. The respective scaling parameters read
$x_c=0.592686$,  $a=0.748386$, and  $b=0.104841$. However, there are no errorbars, yet.
They will be determined next.

\subsection{Estimate errors of the scaling parameters}
\label{ASAsect03subsect02}
\myProg{} estimates the errorbars for the scaling parameters above by means of a so called
$S+1$ analysis, as described in \cite{ASAhoudayer2004}.
Therefore, \myProg{} simply needs to be called with the additional command line parameter
\verb'-getError' as

\begin{lstlisting}
python autoScale.py -f inputFiles.dat -o error.out \
              -xc 0.5927 -a 0.75 -b 0.104 -xr -1.0 1.0 -getError
\end{lstlisting}
This results in the output file \verb'error.out':

\begin{lstlisting}
# S+1 error analysis yields:
# Scaling analysis restricted to
  xr = [-1.000000 : 1.000000]
# <scalePar>  <-Err>  <+Err>
  xc = 0.592686 0.000075 0.000075
   a = 0.748386 0.005913 0.007462
   b = 0.104841 0.000975 0.000975
\end{lstlisting}
Since the calculation of the errors requires some additional effort, it should not be 
performed during repeated calls of \myProg{}, as in subsection \ref{ASAsect03subsect02} above. 
However, once the optimal parameters are identified, the option \verb'-getError' can be used 
to get a grip on the errorbars of the scaling parameters.
By convention, the larger of the two values \verb|<+/-Err>| is taken as the final error bar 
for the respective scaling parameter.

\subsection{Illustration of the data collapse}
\label{ASAsect03subsect03}
%
\begin{figure}[t!]
\begin{center}
\includegraphics[width=1.0\linewidth]{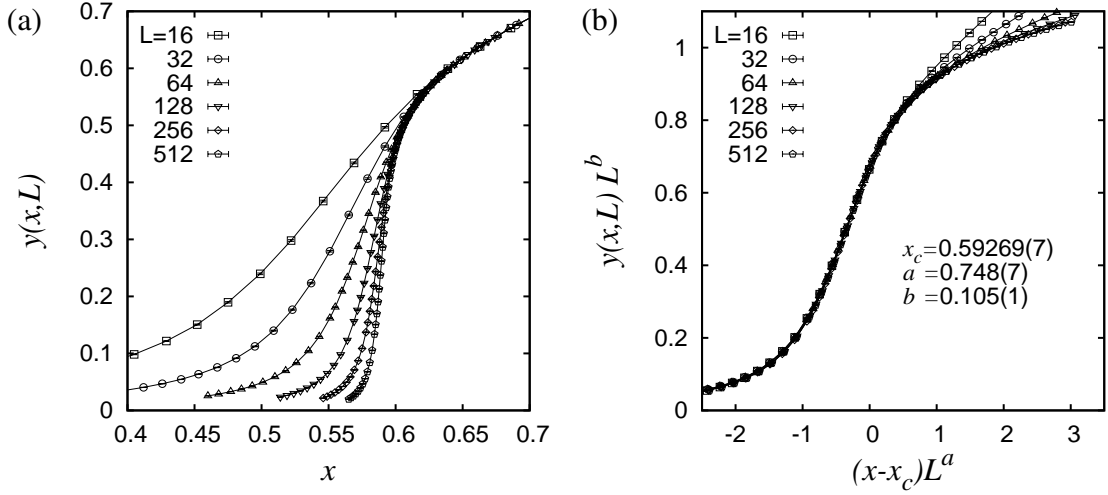}
\caption{\footnotesize
Illustration of the data collapse technique for the order parameter of the site percolation 
problem in $2d$.  
(a) unscaled data for different system sizes $L$ close to the critical 
point. Each data point represents an average over $\approx\!12\,800$
individual occupied/empty configurations simulated at the respective values of $x$ and $L$.
(b) data collapse after the scaling parameter optimization using \myProg{} according 
to the scaling assumption Eq.\ \ref{eq:scalingAssumption}.
\label{fig:scalingPlot}}
\end{center}
\end{figure}
Here, the result of the FSS analysis is illustrated for
the order parameter of the percolation model, as introduced in the preface to the present section.
In this regard, Fig.\ \ref{fig:scalingPlot}(a) shows the unscaled data and Fig.\ \ref{fig:scalingPlot}(b)
illustrates the data collapse obtained for the optimized scaling parameters $x_c$, $a$ and $b$. 
In Figs.\ \ref{fig:scalingPlot}(a),(b), each data point represents an average over $\approx\!12\,800$
individual occupied/empty configurations, simulated at the respective values of $x$ and $L$.
With increasing system size, the critical region narrows as $\sim\!L^{-a}$, see Fig.\ \ref{fig:scalingPlot}(a). 
If the data curves for different system sizes are superposed and rescaled in a proper 
manner, they fall on top of each other, as evident from Fig.\ \ref{fig:scalingPlot}(b).
If one settles for an approximation of the scaling function up to lowest order in 
the respective argument, as mentioned above in section \ref{ASAsect01}, 
one can expect that the scaling assumption holds best close to the critical point and 
that there are corrections to the scaling behavior outside the critical scaling window. 
This can also be seen in Fig.\ \ref{fig:scalingPlot}(b),
where, for values $(x-x_c)L^{a} > 1$ on the rescaled abscissa, the 
curves belonging to different system sizes $L$ start to deviate from each other.
Finally, note that the values of the optimal scaling parameters determined in 
subsections \ref{ASAsect03subsect02} and \ref{ASAsect03subsect03} are in agreement with the
exact exponents for $2d$ percolation (given in the squared braces):
\begin{eqnarray*}
x_c&=&0.59269(7)~[0.59274621(13)] \\
a&=&0.748(7)~[3/4=0.75]\\
b&=&0.105(1)~[15/144\approx0.1042]
\end{eqnarray*}
For the highly precise estimate of the numerical value for $x_c$ in the squared braces see Ref.\ \cite{ASAnewman2000}.

\section{Limitations and tips}
\label{ASAsect04}
Within the presented section, 
subsection \ref{ASAsect04subsect00} explains some requirements that have
to be fulfilled in order to use \myProg{},
subsection \ref{ASAsect04subsect01} discusses the limitations of \myProg{}
and \ref{ASAsect04subsect02} explains how to obtain an initial guess for 
the scaling parameters using the {\tt gnuplot} data plotting tool \cite{ASAgnuplotRef}.
Finally, subsection \ref{ASAsect04subsect03} gives some practical tips on how to 
improve the outcome of a scaling analysis.

\subsection{System requirements}
\label{ASAsect04subsect00}
In order to use \myProg{}, the following requirements have to be fulfilled:
\begin{itemize}
\item Download the source files of this preprint by choosing 
      the download-option \verb|other|. 
      Save the respective source-file archive on your system and extract it in 
      order to find the supplementary archive \verb|SCALING_ANALYSIS.tar.gz|. 
      Extract the supplementary archive via \verb|tar -xzvf SCALING_ANALYSIS.tar.gz|
      and see the readme-file contained therein.
\item \myProg{} has been under development since 2007. It was implemented 
      under {\tt Python} 2.3.4 (and later versions) and it imports only basic 
      {\tt Python}-modules that come with any python installation. 
      Accordingly, check that
      you have installed {\tt Python} 2.3 or a later version on your system. You can
      check your current {\tt Python} version by typing \verb|python -V|. 
      If necessary, see \cite{ASApyRef} or ask your system administrator.
\item Read this users guide and the readme-file contained in the supplementary 
      archive \verb|SCALING_ANALYSIS.tar.gz|.
\end{itemize}

\subsection{Limitations of \myProg{}}
\label{ASAsect04subsect01}
Unfortunately, \myProg{} is not guaranteed to find an optimal set of scaling parameters at all. 
This is due to the drawback that multidimensional minimization algorithms, as the downhill simplex algorithm, 
are likely spoiled by the immediate payoff arising from local minima in the fitness landscape. 
Once caught by such a local minimum, it is hard to escape from it and the scaling parameters are likely to 
converge to non-optimal values.
A remedy for this would be to restart the scaling analysis for a different ``guess'' of the initial scaling parameters
and to check the quality of the resulting scaling parameters visually by using data 
plotting tools as, e.g., {\tt gnuplot} \cite{ASAgnuplotRef}.
However, if the initial scaling parameters are reasonably close to the optimal parameters, the 
minimization procedure will work well.

\subsection{Scaling parameters: how to obtain an initial guess}
\label{ASAsect04subsect02}
As pointed out above, the quality of the initial scaling parameters is crucial for 
a successful finite-size scaling analysis.
In this regard, one can already obtain estimates for the scaling parameters that are
sufficiently close to the optimum in order to ensure a good performance of the
FSS routine by means of simple data plotting tools. 
Unfortunately, the {\tt gnuplot} program itself cannot process command line arguments.
This makes it very inconvenient to tune the scaling parameters by hand.
However, the following shell script is a workaround that can be used to create a 
scaling plot and to visually inspect the data collapse for a given set of scaling parameters
that are passed as command line arguments:

\begin{lstlisting}
# FILE:  scalingPlot.sh
# TASK:  create scaling plot to inspect data collapse
# USAGE: bash scalingPlot.sh <xc> <a> <b>
P='./ORDER_PARAMETER' 	# path to input files
gnuplot -persist << EOF

set key samplen 1. left					# customize key
set xl "(x-xc) L^a"; set yl "y L^b"			# set x,y labels 
xc=$1; a=$2; b=$3					# set scaling parameters
set label 1 "xc=$1\na =$2\nb =$3" at graph 0.7,0.2 	# list scaling parameters
sx(x,L)=(x-xc)*L**a; sy(y,L)=y*L**b			# def scaling assumption

p "$P/orderParam_L16.dat"  u (sx(\$1,16)): (sy(\$2,16))  w lp t "L=16"\
, "$P/orderParam_L32.dat"  u (sx(\$1,32)): (sy(\$2,32))  w lp t "  32"\
, "$P/orderParam_L64.dat"  u (sx(\$1,64)): (sy(\$2,64))  w lp t "  64"\
, "$P/orderParam_L128.dat" u (sx(\$1,128)):(sy(\$2,128)) w lp t " 128"\
, "$P/orderParam_L256.dat" u (sx(\$1,256)):(sy(\$2,256)) w lp t " 256"\
, "$P/orderParam_L512.dat" u (sx(\$1,512)):(sy(\$2,512)) w lp t " 512"
EOF
\end{lstlisting}
This script can be used to obtain a rough guess for the scaling parameters that enter
the scaling assumption. Consequently, \myProg{} can be used to optimize these initial
parameters.

\subsection{Scaling analysis: how to improve the result}
\label{ASAsect04subsect03}
There are several options to improve the results of a scaling analysis:
\begin{itemize}
\item Leave out small system sizes. This will reduce systematic errors that are 
due to finite-size effects.
\item Use at least three data sets. This will ensure that the master function,
computed by \myProg{} in order to determine the quality of the data collapse, behaves
quite smooth.
\item Increase the number of interpolation points in the critical region. This
will significantly improve the master function.
\item Increase the number of samples that make up the individual data points. 
This will improve the statistics of the raw data and also lead to a somewhat
smaller error of the optimal scaling parameters.
\end{itemize}

\section*{Acknowledgments}
OM would like to thank A.\ K.\ Hartmann for critical reading the 
manuscript and I.\ Ilnicki for working through the example once. 
Finally, OM acknowledges financial support from the VolkswagenStiftung
within the program  ``Nachwuchsgruppen an Universit\"aten''. The
simulations were performed at the GOLEM I cluster for scientific 
computing at the University of Oldenburg (Germany).

\renewcommand*\bibname{References}